\documentclass[preprint,12pt,authoryear]{elsarticle}

\usepackage{amssymb}
\usepackage{amsmath}

\usepackage{hyperref}

\newcommand{\Lw}{L_\mathrm{w}}
\newcommand{\uw}{u_\mathrm{w}}
\newcommand{\Mdotw}{\dot{M}_\mathrm{w}}
\newcommand{\tage}{t_\mathrm{age}}

\journal{Journal of High Energy Astrophysics}

\begin{document}

\begin{frontmatter}



\title{On the hadronic origin of the very high energy $\gamma$-ray emission surrounding the young massive stellar cluster Westerlund 1} 


\author[a,b]{Zhaodong Shi} 
\author[a,b,c]{Rui-zhi Yang}

\affiliation[a]{organization={CAS Key Laboratory for Research in Galaxies and Cosmology, Department of Astronomy, School of Physical Sciences,
University of Science and Technology of China},
            city={Hefei},
            postcode={230026}, 
            state={Anhui},
            country={China}}

\affiliation[b]{organization={School of Astronomy and Space Science, University of Science and Technology of China},
            city={Hefei},
            postcode={230026}, 
            state={Anhui},
            country={China}}

\affiliation[c]{organization={TIANFU Cosmic Ray Research Center},
            city={Chengdu},
            state={Sichuan},
            country={China}}

\begin{abstract}
The Westerlund 1 (Wd 1) is the most massive known young star cluster in the Galaxy, and an extended $\gamma$-ray source HESS J1646-458 surrounding it has been detected up to 80 TeV in the very high energy, implying that cosmic rays (CRs) are accelerated effectively in the region. However, the dominant radiation process contributing to the $\gamma$-ray emission is not well constrained. In the present work, we develop a model of CR acceleration at the termination shock in the superbubble inflated by the interaction of the cluster wind from the Wd 1 with the surrounding interstellar medium. We then calculate the flux and radial profile of $\gamma$ rays produced by the inelastic collisions of the hadronic CRs with the ambient gas. Our results with reasonable parameters can explain well the spectrum and radial profile of the $\gamma$-ray emission of HESS J1646-458, and consequently the $\gamma$-ray emission of HESS J1646-458 is likely to be of hadronic origin.
\end{abstract}




\end{frontmatter}



\section{Introduction}
\label{sec:intro}

The origin of cosmic rays (CRs) is still elusive more than one century after their discovery. The ``standard paradigm'' supposes that the Galatic CRs are accelerated at shock fronts generated by supernova explosions \citep{Blasi2013}. Recent $\gamma$-ray observations of supernove remnants \citep[SNRs;][]{Funk2015}, however, imply that it is troubling for SNRs to accelerate CRs up to the ``knee'' ($\sim 3$ PeV), thus the alternative scenarios, such as the young massive clusters (YMCs), have been proposed in the context \citep[e.g.][]{Aharonian2019, Vieu2023, Peron2024}. It has been proposed long before that the CRs can be accelerated by the massive stellar winds and in superbubbles (SBs) generated by the collective interactions of massive stars \citep{Cesarsky1983, Parizot2004}. Moreover, the anomalous excess of the isotope abundance ratio $^{22}\mathrm{Ne}/^{20}\mathrm{Ne}$ observed in the Galactic CRs suggests that SBs are the likely source of at least a substantial fraction of Galactic CRs \citep{Binns2005}.

In the last two decades, it has been well established that YMCs and SBs are among the high energy $\gamma$-ray sources, including Westerlund 1 \citep[Wd 1;][]{HC2012, HC2022}, Westerlund 2 \citep{Yang2018}, RCW 38 \citep{Ge2024, Pandey2024}, Cygnus bubble \citep{Grenier2011, Blandford2021, Cao2024} in our Galaxy, 30 Dor C \citep{Komin2015}, and R136 in the Large Magellanic Cloud \citep{HESS2024}. Therefore, YMCs are viable to be the accelerators of Galactic CRs and the potential PeVatrons accelerating CRs up to the ``knee'', and in fact, it has been demonstrated recently by LHAASO that the Cygnus bubble is a Galactic PeVatron \citep{Cao2024}. In the context, the interests in YMCs and SBs on their contribution to the origin of CRs renew. However, it is not well understood how CRs are accelerated by YMCs and where their acceleration sites are in SBs \citep{Gupta2018, Bykov2020, Vieu2022b}. Furthermore, the dominant process contributing to the production of very high energy $\gamma$ rays can not be pinned down in many cases, and both hadronic and leptonic scenarios can explain the observations. In the hadronic scenario, $\gamma$ rays originate from the decay of neutral $\pi^0$ mesons which are generated from the inelastic collisions of hadronic CRs with interstellar gas, while in the leptonic one, $\gamma$ rays are produced through the inverse Compton (IC) scattering of CR electrons on the low energy seed photons.

The Wd 1 is the most massive known YMC in our Galaxy, which hosts many evolved massive stars \citep[see][]{Zwart2010}. An extended very high energy $\gamma$-ray source HESS J1646-458 surrounding the Wd 1 is detected by H.E.S.S. \citep{HC2012} and investigated in detail \citep{HC2022}. A prominent feature in the morphology of HESS J1646-458 is that centering on the Wd 1 the $\gamma$-ray emission exhibits a shell-like structure, which has significant implication on the processes dominating the acceleration of CRs and the production of $\gamma$-ray radiation. The termination shock (TS) of cluster wind forming around the Wd 1 has been proposed to be the site accelerating CRs, since the radial distance to the Wd 1 of the peak in the $\gamma$-ray excess profile is consistent with the theoretical estimate on the radius of the TS. However, it is still not clear whether the $\gamma$-ray emission from HESS J1646-458 is of hadronic or leptonic origin. \citet{HC2022} do not find a clear correlation of the $\gamma$-ray emission with the gas distribution traced by HI and CO, while \citet{Haerer2023} argue recently that the $\gamma$-ray emission of HESS J1646-458 is of leptonic origin. In the context, we examine the hadronic scenario for the $\gamma$-ray emission of HESS J1646-458 in more detail.

The acceleration of CRs at the TS of a stellar wind has been studied long before by \citet{Webb1985}. Recently, the acceleration of CRs at the TS resulting from the interaction of the collective wind of star clusters with the surrounding interstellar medium (ISM) has been investigated by \citet{Morlino2021} and \citet{Blasi2023}. In the present article, we address the scenario in which the CRs (mainly protons) are accelerated at the TS of the cluster wind from the Wd 1. The organization of the article is as follows. We first describe the structure of the SB forming around the Wd1 in Section \ref{sec:sb}. We then develop the model of CR acceleration at the TS in Section \ref{sec:accel}, and in Section \ref{sec:gam}, we calculate the spectrum and radial profile of $\gamma$ rays resulting from the inelastic collisons of hadronic CRs with the ambient gas and compare with those of the $\gamma$-ray emission of HESS J1646-458. Finally, we conclude and discuss the implications of our results  in Section \ref{sec:conclusion}.

\section{Superbubble surrounding the Wd 1}
\label{sec:sb}

The Wd 1 is compact YMC with a half-mass radius of about 1 pc. The cluster wind from the Wd 1 impacts the surrounding ISM and inflates a cavity, i.e., a SB. The analytic theory of wind-blown bubbles is developed by \citet{Castor1975} and \citet{Weaver1977}.

\begin{figure}
    \centering
    \includegraphics[width=0.5\linewidth]{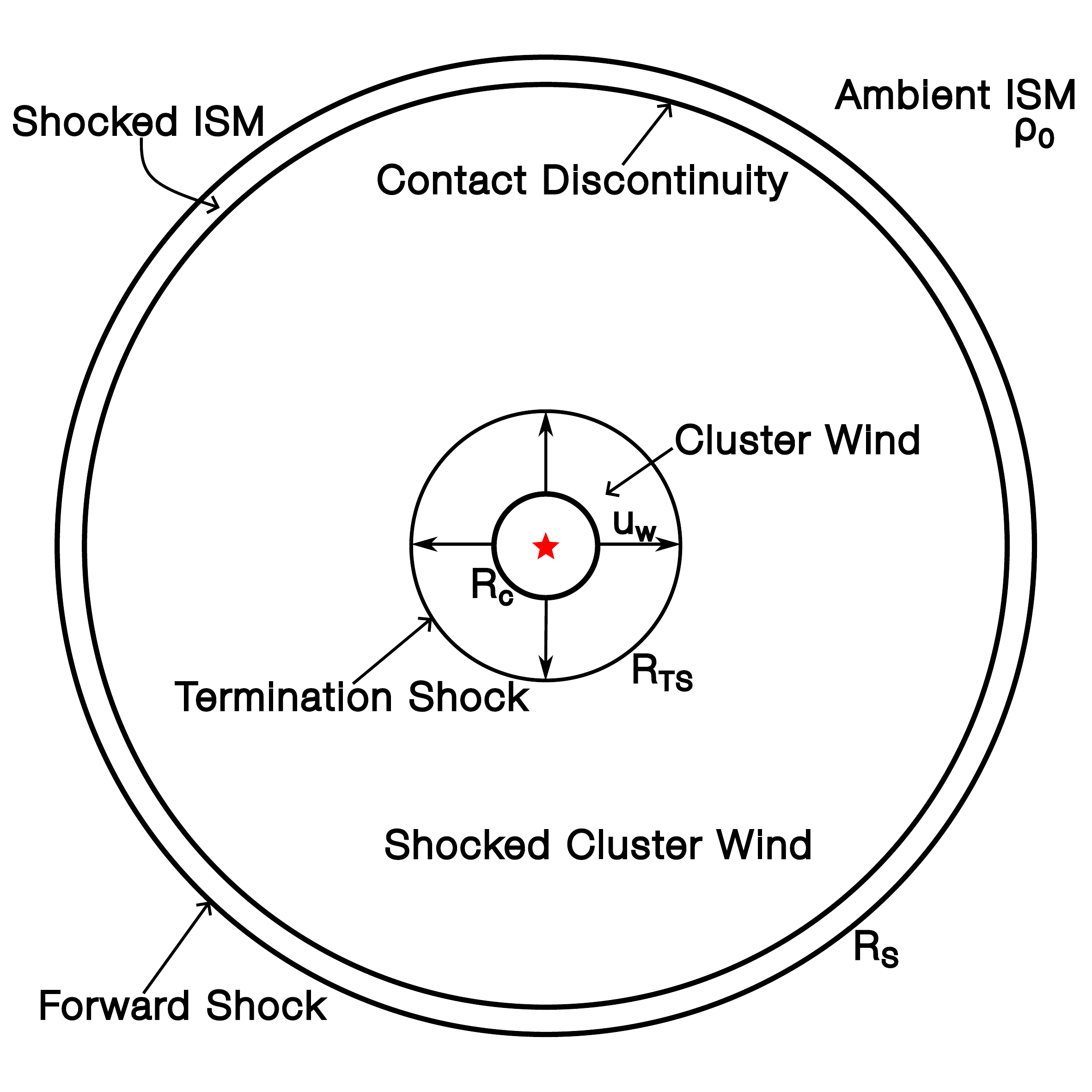}
    \caption{Schematic structure of a superbubble inflated by the cluster wind of a massive star cluster.}
    \label{fig:sbstructure}
\end{figure}

Figure \ref{fig:sbstructure} is an illustration which depicts schematically the structure of a SB. The SB is separated from the surrounding unshocked ISM by a supershell, which consists of the shocked ISM. The supershell is separated from the subsonic cavity, which consists of shocked cluster wind, by a contact discontinuity. For the Wd 1 whose age $\tage = 4~\mathrm{Myr}$ \citep{Clark2005, Brandner2008, HC2022}, the supershell has collapsed and is in fact thin due to cooling. Assuming that the kinetic power of the cluster wind is $\Lw = \frac{1}{2}\Mdotw \uw^2$, where $\Mdotw$ is the mass loss rate and $\uw$ is the wind speed, the radius of the supershell is

\begin{equation} \label{eq:shell}
  \begin{aligned}
    R_S &= \left(\frac{125}{154\pi} \frac{\xi_b \Lw \tage^3}{\rho_0}\right)^{1/5} \\ &= 70.45~\mathrm{pc}~\left(\frac{\xi_b \Lw}{10^{38}~\mathrm{erg/s}}\right)^{1/5} \\ &\times \left(\frac{\rho_0}{1~m_p~\mathrm{cm^{-3}}}\right)^{-1/5} \left(\frac{\tage}{1~\mathrm{Myr}}\right)^{3/5},
  \end{aligned}
\end{equation}

\noindent and the swept-up mass of the supershell is

\begin{equation} \label{eq:mass}
  \begin{aligned}
    M_S &= \frac{4\pi}{3}\rho_0 \left(\frac{125}{154\pi} \frac{\xi_b \Lw \tage^3}{\rho_0}\right)^{3/5} \\ &= 3.62\times10^4~M_\odot~\left(\frac{\xi_b \Lw}{10^{38}~\mathrm{erg/s}}\right)^{3/5} \\ &\times \left(\frac{\rho_0}{1~m_p~\mathrm{cm^{-3}}}\right)^{2/5} \left(\frac{\tage}{1~\mathrm{Myr}}\right)^{9/5},
  \end{aligned}
\end{equation}

\noindent where $\rho_0$ is the mass density of external ISM, $m_p$ is the proton mass, $M_\odot$ is the solar mass, and $\xi_b$ is the fraction of $\Lw$ converted to inflate the SB. The factor $\xi_b$ is introduced by \citet{Vieu2022a} to account for the so-called energy crisis of SBs \citep{Cooper2004, Kavanagh2020}, and they estimate $\xi_b \approx 0.22$. For the Wd 1, its kinetic luminosity $\Lw = 10^{39}$ erg/s and the speed of its cluster wind $\uw=2500$ km/s \citep{Haerer2023}.

The shocked wind cavity is isobaric, and its pressure is 

\begin{equation} \label{eq:pressure}
  \begin{aligned}
    P &= \frac{7}{25}\rho_0 \left(\frac{125}{154\pi} \frac{\xi_b \Lw}{\rho_0 \tage^2}\right)^{2/5} \\ &= 2.22 \times 10^{-11}~\mathrm{dyne/cm^2}~\left(\frac{\xi_b \Lw}{10^{38}~\mathrm{erg/s}}\right)^{2/5} \\ &\times \left(\frac{\rho_0}{1~m_p~\mathrm{cm^{-3}}}\right)^{3/5} \left(\frac{\tage}{1~\mathrm{Myr}}\right)^{-4/5}.
  \end{aligned}
\end{equation}

\noindent The shocked wind cavity is separated from the unshocked supersonic cluster wind by a termination shock (TS). Across the TS and assuming that the TS is strong with a compression ratio of 4, the pressure immediately downstream of the TS according to Rankine-Hugoniot conditions is 

\begin{equation} \label{eq:jump}
P(R_\mathrm{TS}^+) = \frac{3}{4}\frac{\Mdotw \uw}{4\pi R_\mathrm{TS}^2},
\end{equation}

\noindent where $R_\mathrm{TS}$ is the radius of the TS. By equating Eq. \eqref{eq:pressure} to Eq. \eqref{eq:jump}, we get

\begin{equation}
  \begin{aligned}
    R_\mathrm{TS} &= \sqrt{\frac{3 (3850\pi)^{2/5}}{56\pi}} \left(\frac{\Lw}{\rho_0}\right)^{3/10} \uw^{-1/2} \tage^{2/5} \xi_b^{-1/5} \\
    &= 23.75~\mathrm{pc}~\left(\frac{\Lw}{10^{38}~\mathrm{erg/s}}\right)^{3/10} \left(\frac{\rho_0}{1~m_p~\mathrm{cm^{-3}}}\right)^{-3/10} \\ &\times \left(\frac{\uw}{1000~\mathrm{km/s}}\right)^{-1/2} \left(\frac{\tage}{1~\mathrm{Myr}}\right)^{2/5} \xi_b^{-1/5}.
  \end{aligned}
\end{equation}

Finally, the mass density in the shocked wind cavity \citep{MacLow1988, Vieu2022a} is 

\begin{equation} \label{eq:rho-cavity}
  \begin{aligned}
    \rho &= 0.019~m_p~\mathrm{cm}^{-3}~\left(\frac{\xi_b \Lw}{10^{38}~\mathrm{erg/s}}\right)^{6/35} \\ &\times \left(\frac{\rho_0}{1~m_p ~\mathrm{cm}^{-3}}\right)^{19/35} \left(\frac{\tage}{1~\mathrm{Myr}}\right)^{-22/35}.
  \end{aligned}
\end{equation}

\noindent The above expression assumes that there is no any dense clumpy gas residing in the shocked wind cavity, which escapes from being swept up by the forward shock.

\section{CR acceleration at the TS and transport within The SB} \label{sec:accel}

The acceleration of CRs at the TS and their transport inside the SB is described by the following well-known equation in spherical symmetry \citep[e.g.][]{Lopez2004,Morlino2021}

\begin{equation} \label{eq:parker}
\frac{\partial f}{\partial t} = \frac{1}{r^2}\frac{\partial }{\partial r}\left(r^2D \frac{\partial f}{\partial r}\right) - u\frac{\partial f}{\partial r} + \frac{1}{3r^2}\frac{\partial r^2 u}{\partial r}p\frac{\partial f}{\partial p} + Q,
\end{equation}

\noindent where $f(r, p)$ is the distribution function of CRs, $D(p)$ is the spatial diffusion coefficient, and $u(r)$ is the advection velocity, and $Q(r, p)$ is the source term. We shall assume that the steady-state solution, namely $\partial f/\partial t=0$, is reached. We have also solved numerically the time-dependent transport equation, and comparing the steady-state solutions with the time-dependent ones, we have verified that the steady-state assumption is valid for the cases we are investigating. In the present article, we do not address the stochastic acceleration due to magnetic turbulence inside SBs \citep{Tolksdorf2019, Vieu2022a}.

Since the nature of magnetic turbulence inside SBs is not well known, we assume a homogeneous spatial diffusion coefficient phenomenologically within the SB, and that the upstream diffusion coefficient is the same as the downstream one, namely

\begin{equation} \label{eq:diffcoeff}
D(p) = D_0 \beta \left(\frac{p}{1~\mathrm{GeV/c}}\right)^\delta,
\end{equation}

\noindent where $\beta$ is the ratio of CR velocity to the light speed $c$.

The advection velocity is 

\begin{equation} \label{eq:advvel}
u(r) = 
\begin{cases}
\uw, & r < R_\mathrm{TS},\\
\frac{\uw}{4}\frac{R_\mathrm{TS}^2}{r^2}, & r > R_\mathrm{TS}.
\end{cases}
\end{equation}

\noindent There is a jump in the advection velocity according to Rankine-Hugoniot conditions across the TS, and the flow downstream of the TS is incompressible \citep{Weaver1977}.

We assume that monoenergetic CRs are injected continually at the TS, thus the source term is

\begin{equation}
Q(r,p)= \frac{\eta_\mathrm{inj} \Lw}{E_\mathrm{k,inj}} \frac{\delta(p - p_\mathrm{inj})}{4\pi p_{\mathrm{inj}}^2} \frac{\delta(r - R_\mathrm{TS})}{4\pi R_{\mathrm{TS}}^2},
\end{equation}

\noindent where $\eta_\mathrm{inj}$ is the injection efficiency, $p_\mathrm{inj}=1~\mathrm{GeV/c}$ is the injection momentum, and the injection kinetic energy $E_\mathrm{k,inj} = \sqrt{p_\mathrm{inj}^2c^2 + m_p^2c^4} - m_pc^2$.

For solving Eq. \eqref{eq:parker}, we need specify the boundary conditions. We assume a reflection boundary condition, $\partial{f}/\partial{r}$ = 0, at the inner radial boundary which is at $R_c$ = 1 pc, and an absorbing boundary condition, $f=0$, at the outer radial boundary $R_S$ \citep{Florinski2003}. For the latter, \citet{Menchiari2024} adopted a flux conservation boundary condition at $r=R_S$, and we have verified that there is only little difference between results obtained by adopting the two different boundary conditions, given that the diffusion coefficient $D_{0,\mathrm{ISM}}$ in the external ISM takes the canonical value, namely $D_\mathrm{0,ISM}\approx3\times10^{28}~\mathrm{cm^2/s}$ at about 1 GeV \citep{Strong2007}.

Eq. \eqref{eq:parker} can be solved analytically downstream of the TS, due to $\partial(r^2u)/\partial r$=0, and the solution for $R_\mathrm{TS} < r <R_S$ is

\begin{equation}
f(r, p) = f_\mathrm{TS}(p)\frac{1 - \exp\left[-\frac{\uw R_\mathrm{TS}^2}{4D(p)}(1 / r - 1 / R_S)\right]}{1 - \exp\left[-\frac{\uw R_\mathrm{TS}}{4D(p)}(1 - R_\mathrm{TS}/R_S)\right]},
\end{equation}

\noindent where $f_\mathrm{TS}(p)$ is the CR distribution function at the TS, which cannot be determined at present \citep{Morlino2021}.

Across the TS, we require the continuity condition $f(R_\mathrm{TS}^+, p) = f(R_\mathrm{TS}^-, p) = f_\mathrm{TS}(p)$, and the matching condition \citep[see][]{Potgieter1988, Morlino2021} as following

\begin{equation} \label{eq:matching}
\left.\left(D\frac{\partial f}{\partial r}\right)\right|_1^2 + \frac{u_2 - u_1}{3}p\frac{\partial f_\mathrm{TS}(p)}{\partial p} + \frac{1}{4\pi R_\mathrm{TS}^2}\frac{\eta_\mathrm{inj} \Lw}{E_\mathrm{k,inj}}\frac{\delta(p-p_\mathrm{inj})}{4\pi p_\mathrm{inj}^2} = 0,
\end{equation}

\noindent where the labels 1 and 2 denote the immediately upstream and downstream of the TS, respectively, and $u_1 = \uw$ and $u_2 = \uw/4$. Eq. \eqref{eq:parker} is solved numerically upstream of the TS via finite difference methods, while Eq. \eqref{eq:matching} is regarded as the boundary condition at the TS. The numerical methods adopted are Crank-Nicolson difference scheme for the spatial diffusion and semi-implicit upwind difference scheme for the advection \citep{nr1992}. In addition, the explicit difference scheme is applied for the momentum loss (the wind upstream of the TS is expanding, hence CRs will loss energy). The transport of CRs upstream of the TS is similar to that of CRs in the solar wind \citep[namely solar modulation; see][]{leRoux1996, Lopez2004}.

\begin{figure}
    \centering
    \includegraphics[width=0.5\linewidth]{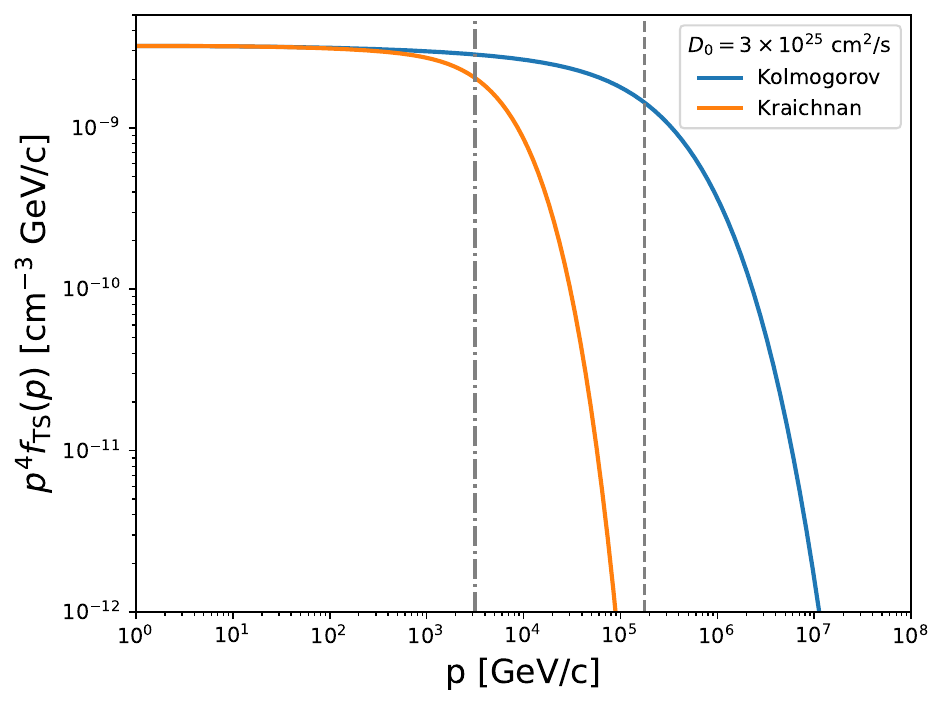}\\
    \includegraphics[width=0.5\linewidth]{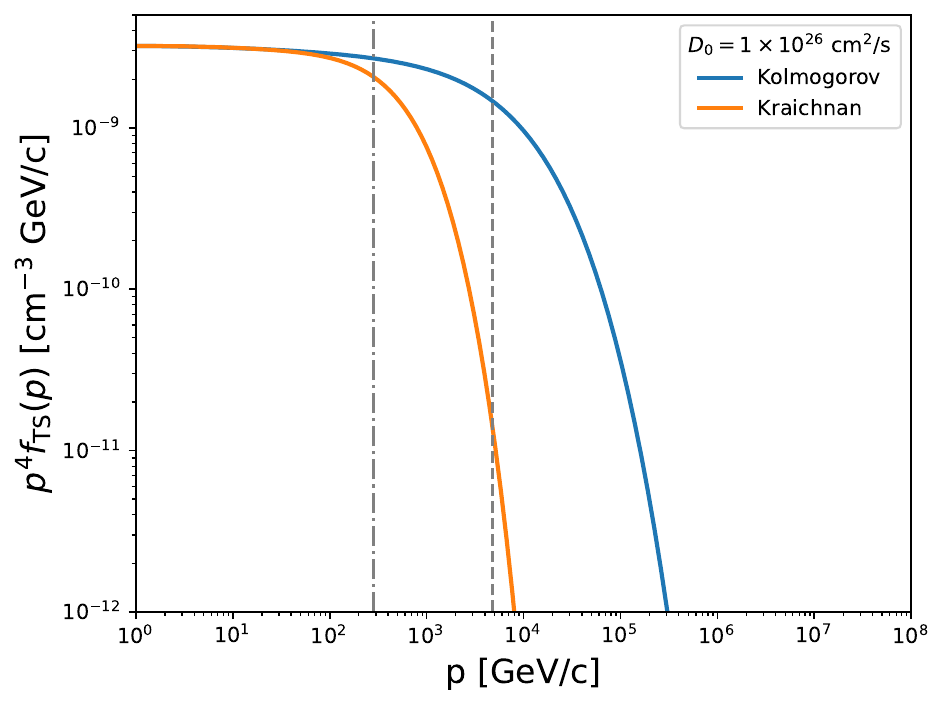}
    \caption{\textbf{Top}: The distribution functions of accelerated CRs at the TS for Kolmogorov ($\delta=1/3$) and Iroshnikov-Kraichnan ($\delta=1/2$) turbulence spectra. The vertical dashed and dotdashed lines show the cutoff momenta of accelerated CRs as defined by Eq. \eqref{eq:pcut} for Kolmogorov and Iroshnikov-Kraichnan turbulence spectra, respectively. The injection efficiency $\eta_\mathrm{inj}=0.1$, the spatial diffusion coefficient $D_0 = 3\times 10^{25}~\mathrm{cm^2/s}$, and the mass density of external ISM $\rho_0=50~m_p~\mathrm{cm}^{-3}$,  while other parameters are specified in Section \ref{sec:sb}. \textbf{Bottom}: Same as the top panel but $D_0=1\times 10^{26}~\mathrm{cm^2/s}$.}
    \label{fig:cr-spec-ts}
\end{figure}

Figure \ref{fig:cr-spec-ts} shows the distribution functions of accelerated CRs at the TS for Kolmogorov ($\delta=1/3$) and Iroshnikov-Kraichnan ($\delta=1/2$) turbulence spectra. In the top panel, we have assumed that $D_0=3\times 10^{25}~\mathrm{cm^2/s}$, $\rho_0=50~m_p~\mathrm{cm^{-3}}$, and $\eta_\mathrm{inj}=0.1$, while other parameters have been specified in Section \ref{sec:sb}. Moreover, the bottom panel has the same settings as the top one but $D_0 = 1\times 10^{26}~\mathrm{cm^2/s}$. The TS is effectively planar for the CRs, whose characteristic length $\ell(p) = D(p)/\uw$ is much less than the curvature radius, $R_\mathrm{TS}$, of the TS \citep{leRoux1996}, thus the accelerated spectra have a power-law index of 4 as predicted by the theory of diffusive shock acceleration \citep{Drury1983} for strong shocks. At higher energies, namely when the characteristic length $\ell(p)$ is approaching and larger than $R_\mathrm{TS}$, the accelerated spectra deviate from the power law and cut off. The cutoff momentum of accelerated CRs, $p_c$, can be defined by the relation $\ell(p_c)=\chi R_\mathrm{TS}$ \citep{Peretti2025},  where we should expect $\chi \le 1$. According to our numerical experiments for computing the accelerated spectra by adopting different sets of transport parameters, we observe that $\chi$ = 0.1 is a reasonable value, and the same value is also adopted by \citet{Diesing2023} and \citet{Corso2023} for determining the maximum momentum. Therefore, the cutoff momentum

\begin{equation} \label{eq:pcut}
  \begin{aligned}
   p_c &= \left(\frac{\chi R_\mathrm{TS} u_\mathrm{w}}{D_0}\right)^{1/\delta} \\ &= \left(6.171 \frac{\chi}{0.1} \frac{R_\mathrm{TS}}{20~\mathrm{pc}} \frac{u_\mathrm{w}}{1000~\mathrm{km/s}}\right)^{1/\delta}\left(\frac{D_0}{10^{26}~\mathrm{cm^2/s}}\right)^{-1/\delta}~\mathrm{GeV/c}.
  \end{aligned}
\end{equation}

\noindent The vertical dashed and dotdashed lines in the top panel of Figure \ref{fig:cr-spec-ts} show the cutoff momenta $p_c = 1.77\times 10^5$ GeV/c and $p_c = 3.15\times 10^3$ GeV/c for Kolmogorov and Iroshnikov-Kraichnan turbulence spectra, respectively, provided the parameters we adopt. Likewise, the corresponding cutoff momenta in the bottom panel are $p_c = 4.78\times10^3$ GeV/c and $p_c=284$ GeV/c for Kolmogorov and Iroshnikov-Kraichnan turbulence spectra, respectively.

\begin{figure}
    \centering
    \includegraphics[width=0.5\linewidth]{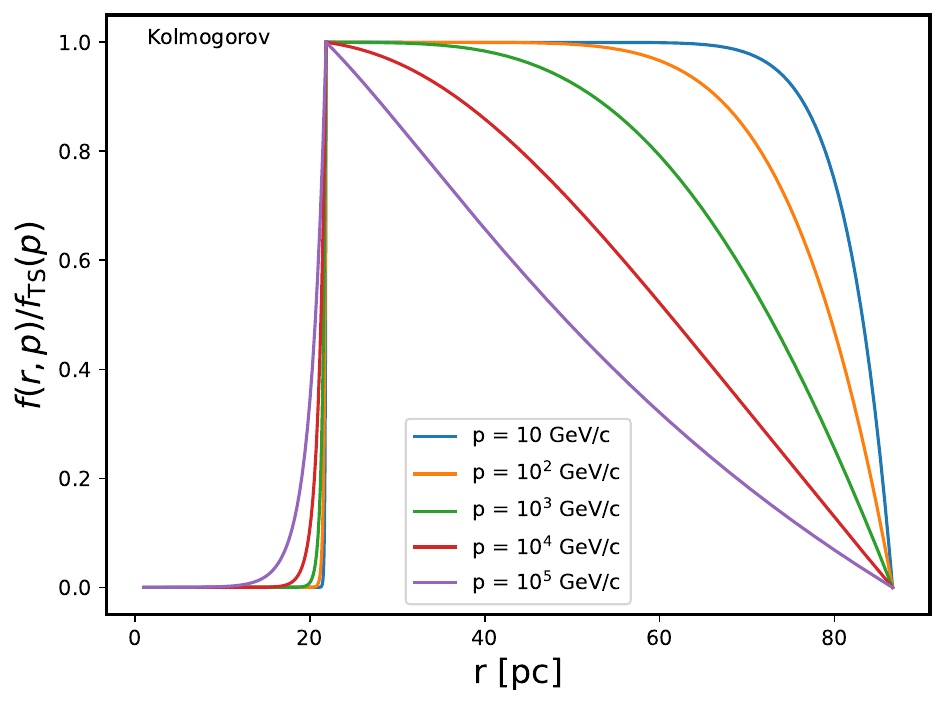} \\
    \includegraphics[width=0.5\linewidth]{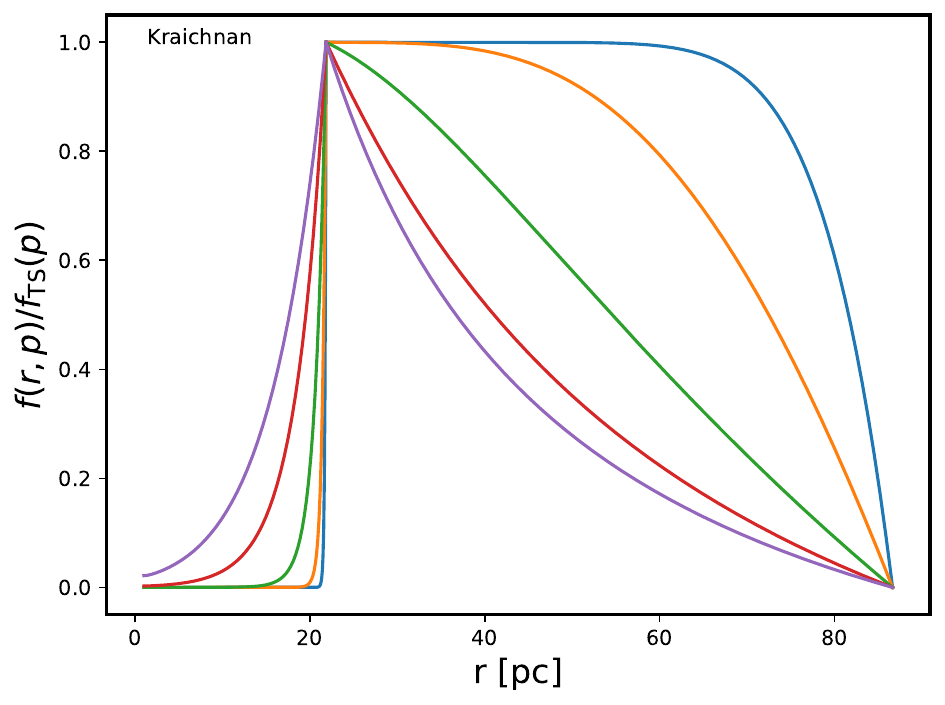}
    \caption{Top and bottom panels show the radial profiles of CR distribution functions at different momenta for Kolmogorov ($\delta=1/3$) and Iroshnikov-Kraichnan ($\delta=1/2$) turbulence spectra, respectively. The injection efficiency $\eta_\mathrm{inj}=0.1$, the spatial diffusion coefficient $D_0 = 3\times 10^{25}~\mathrm{cm^2/s}$, and the mass density of external ISM $\rho_0=50~m_p~\mathrm{cm}^{-3}$,  while other parameters are specified in Section \ref{sec:sb}.}
    \label{fig:cr-profile}
\end{figure}

Figure \ref{fig:cr-profile}, which has the same settings as the top panel of Figure \ref{fig:cr-spec-ts}, shows the radial profiles of CR distriubtion functions at different momenta for Kolmogorov (top panel) and Iroshnikov-Kraichnan (bottom panel) turbulence spectra. It is interesting that the CRs with higher energies concentrate on the narrower region around the TS. Due to the momentum-dependent diffusion, CRs escape from the downstream of the TS more quickly at higher energies since the diffusion dominates over the advection. On the other hand, the CR acceleration time $t_\mathrm{acc} \propto D(p)/\uw^2$ \citep{Drury1983}, and consequently CRs can only be accelerated to higher energies when they can be confined to the neighboring region around the TS for longer time.

\begin{figure}
    \centering
    \includegraphics[width=0.5\linewidth]{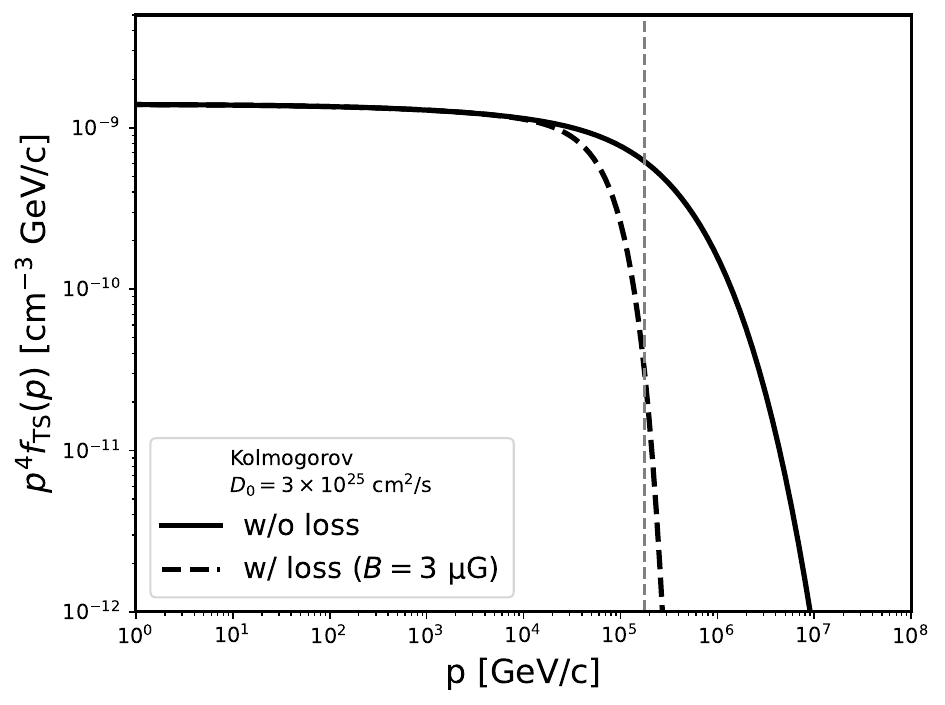} \\
    \includegraphics[width=0.5\linewidth]{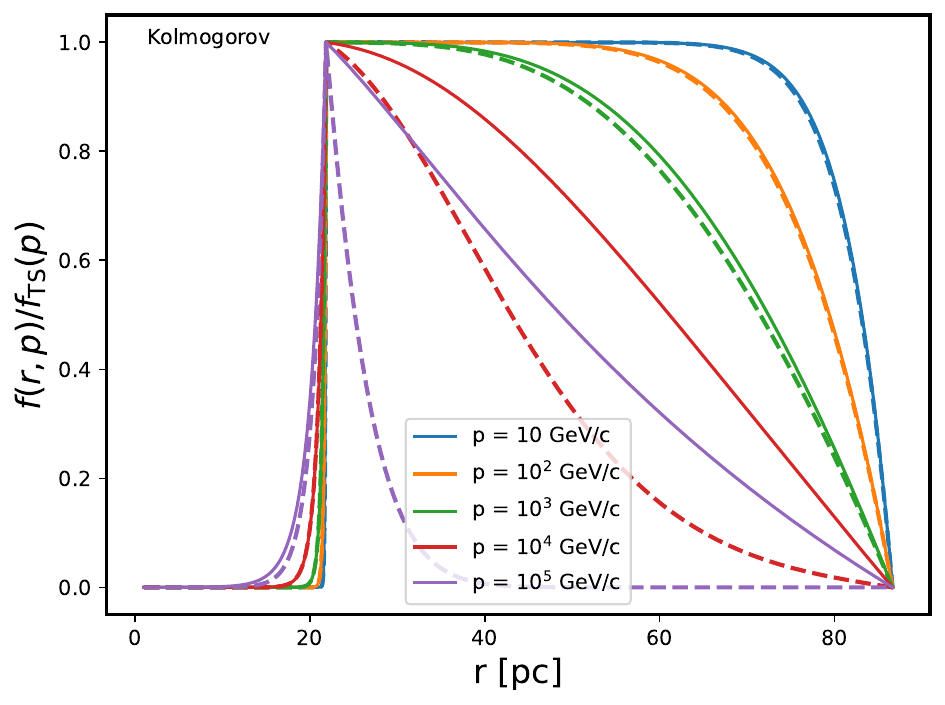}
    \caption{Top and bottom panels show the spectra of accelerated electrons at the TS and their radial profiles at different energies, respectively, for Kolmogorov ($\delta=1/3$) turbulence spectrum. Dashed and solid lines correspond to the cases with and without energy loss resulting from synchrotron radiation taken into account, respectively. The vertical dashed line at the top panel shows the cutoff momentum of accelerated electrons as defined by Eq. \eqref{eq:pcut} when synchrotron cooling is not taken into account. For the case that the energy loss due to synchrotron radiation is taken into account, the magnetic field is $B=3~\mathrm{\mu G}$. The injection efficiency $\eta_\mathrm{inj}=0.1$, the spatial diffusion coefficient $D_0 = 3\times 10^{25}~\mathrm{cm^2/s}$, and the mass density of external ISM $\rho_0=50~m_p~\mathrm{cm}^{-3}$,  while other parameters are specified in Section \ref{sec:sb}.}
    \label{fig:espec}
\end{figure}

Though we are mainly interested in CR protons and the resulted hadronic $\gamma$-ray emission, we discuss briefly the acceleration of CR electrons by the TS. Since electrons suffer a severe energy loss resulting from synchrotron radiation and IC scattering off background soft photons, it is crucial to take the loss into account for investigating their acceleration \citep{Vannoni2009}. We have solved numerically the steady-state transport equation which includes an extra term taking the energy loss due to synchrotron radiation into account, and we have assumed a magnetic field of $B=3~\mathrm{\mu G}$. As a lower limit for the energy loss, we do not take the energy loss due to IC scattering into account. In the top panel of Figure \ref{fig:espec}, the dashed line shows the distribution function at the TS of accelerated electrons when the energy loss due to synchrotron radiation is taken into account, while the solid line shows the corresponding one without synchrotron cooling. As an illustration, we have assumed a Kolmogorov turbulence spectrum, the spatial diffusion coefficient $D_0=3\times 10^{25}~\mathrm{cm^2/s}$, and the mass density of external ISM $\rho_0=50~m_p~\mathrm{cm^{-3}}$. When the synchrotron cooling is not taken into account, electrons can be accelerated up to the maximum momentum $p_c=1.77\times10^5$ GeV/c (vertical gray line) as predicted by Eq. \eqref{eq:pcut}. When synchrotron cooling is taken into account, however, the maximum momentum of electrons is much lower than that given by Eq. \eqref{eq:pcut}. Moreover, in the bottom panel of Figure \ref{fig:espec}, the dashed and solid lines show the radial profiles of accelerated electrons, with and without synchrotron cooling taken into account, respectively. When the synchrotron cooling is taken into account, we can see that the radial distributions of electrons with momentum of 10 TeV/c (red line) and 100 TeV/c (purple line) are much narrower than the corresponding ones without synchrotron cooling.

In the next section, we will calculate the spectra and radial profiles of $\gamma$ rays resulting from the decay of neutral $\pi^0$ mesons generated by the inelastic collisions of CRs with the ambient gas nuclei. Whereas recent hydrodynamic simulations suggest that winds in YMCs can drive Kolmogorov-like turbulence \citep{Gallegos-Garcia2020}, for clarity, we only consider the Kolmogorov turbulence spectrum henceforth, namely $\delta=1/3$ as the fiducial value.

\section{The hadronic $\gamma$ rays} \label{sec:gam}

The emissivity per hydrogen atom of $\gamma$ rays resulting from the inelastic collisions of CR protons with the ambient gas hydrogen \citep{Kafexhiu2014} is 

\begin{equation}
    q_{p\mathrm{H}\to\gamma}(E_{\gamma}) = \int \frac{d\sigma_{p\mathrm{H}\to\gamma}}{dE_{\gamma}}(E_p, E_{\gamma}) I_p(E_p)dE_p,
\end{equation}

\noindent where $E_p$ and $E_\gamma$ are the energies of CR protons and produced $\gamma$-ray photons, respectively, $d\sigma_{p\mathrm{H}\to\gamma}/dE_\gamma(E_p, E_\gamma)$ is the differential production cross sections of $\gamma$ rays, and the CR proton intensity $I_p(E_p) = p^2f(p)$. In the present article, we use the tabulated differential production cross sections of $\gamma$ rays provided by the Python package \texttt{aafragpy} \citep{Koldobskiy2021} to calculate the emissivity.

The $\gamma$-ray flux is 

\begin{equation} \label{eq:gam-flux}
    F_{\gamma}(E_{\gamma}) = \varepsilon_\mathrm{M}\frac{1}{d^2}\int q_{p\mathrm{H}\to\gamma}(E_{\gamma}) n_\mathrm{H} dV,
\end{equation}

\noindent where the integration is over the source volume, $d$ is the distance to the source which is $3.9$ kpc for the Wd 1 \citep[see][]{Kothes2007, Davies2019}, $n_\mathrm{H} = \rho / \mu m_p$ is the ambient gas hydrogen number density assuming that the mean molecular weight is $\mu=1.4$, and $\varepsilon_\mathrm{M}$ is the nuclear enhancement factor accounting for the contribution to the $\gamma$-ray production of nuclei heavier than hydrogen in both CRs and ambient gas, for which we adopt $\varepsilon_\mathrm{M}=2.0$ \citep{Mori2009, Kachelriess2014}. The $\gamma$-ray intensity is

\begin{equation} \label{eq:gam-intensity}
    I_\gamma(E_\gamma) = \varepsilon_\mathrm{M}\int q_{p\mathrm{H}\to\gamma}(E_\gamma)n_\mathrm{H} dl,
\end{equation}

\noindent where the integration is along the line of sight.

\citet{HC2022} find that the $\gamma$-ray flux from HESS J1646-458 can be fitted by a power law with exponential cutoﬀ, and the cutoff energy is about 40 TeV. The $\gamma$-ray photon resulting from the inelastic proton-proton collisions carries averagely about 10 percent of the initial CR proton energy \citep{Kelner2006}, hence we can fix the diffusion coefficient $D_0$ according to Eq. \eqref{eq:pcut} once the external ISM mass density $\rho_0$ is given. We find that adopting $D_0=3\times 10^{25}~\mathrm{cm^2/s}$ can lead to acceptable fits to the $\gamma$-ray flux from HESS J1646-458 for $\rho_0 = $ 30, 40, and 50 $m_p~\mathrm{cm^{-3}}$. Thus, we do not fine-tune the diffusion coefficient $D_0$ in our model fittings, and fix it with the above value. Once both the diffusion coefficient $D_0$ and the mass density of external ISM $\rho_0$ are fixed, the only free parameter is the normalization factor $\eta_\mathrm{inj}\rho_2$, which is the product of injection efficient and the density of shocked wind cavity, and can be constrained by the observed $\gamma$-ray flux. We seek the best-fit normalization factor by using the Python package \texttt{iminuit} \citep{iminuit}. While minimizing the $\chi^2$, we only make use of $\gamma$-ray flux data of HESS J1646-458 to constrain the normalization factor, however, our best-fit results can also explain reasonably the radial profile of $\gamma$-ray intensity of HESS J1646-458.

\begin{figure}
    \centering
    \includegraphics[width=0.5\linewidth]{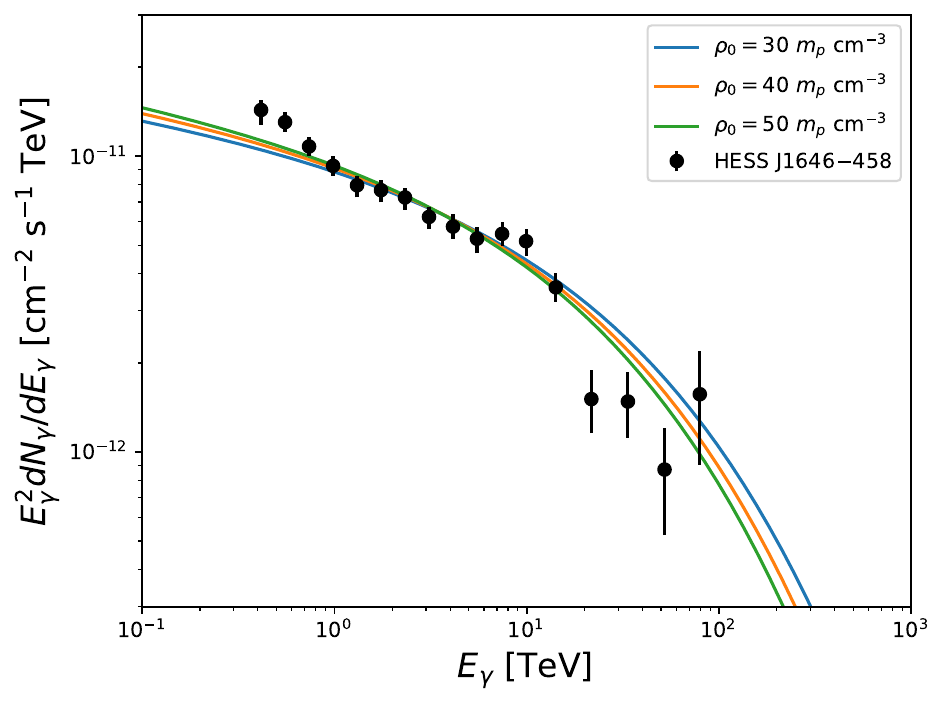}\\
    \includegraphics[width=0.5\linewidth]{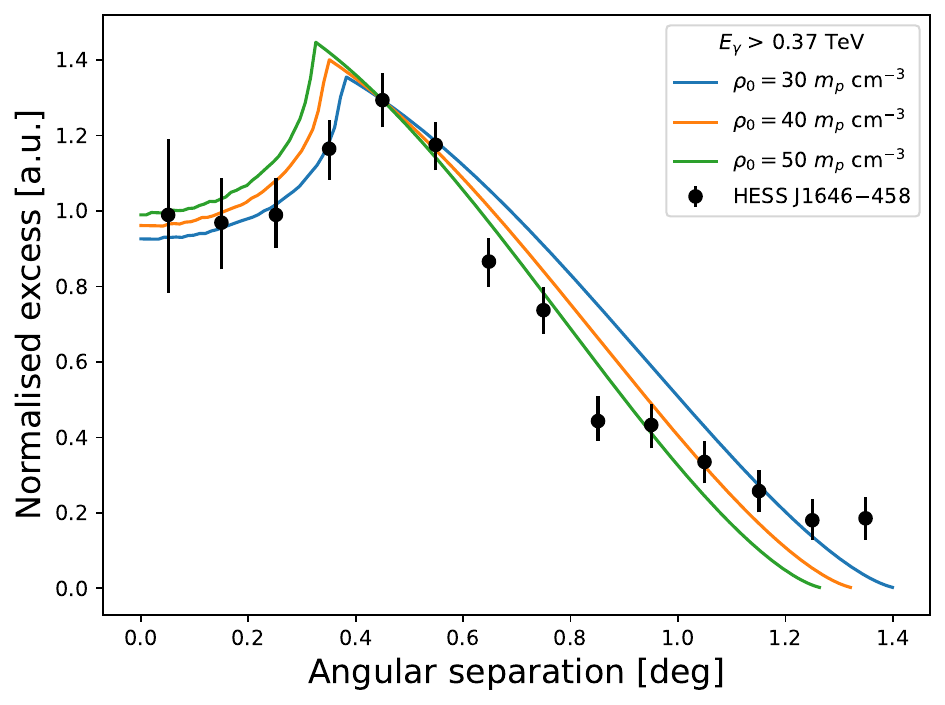}
    \caption{The $\gamma$-ray fluxes (top panel) and the radial profiles of $\gamma$-ray intensities above the threshold of 0.37 TeV (bottom panel) for three different models with the external ISM mass density $\rho_0 =$ 30, 40, and 50 $m_p~\mathrm{cm^{-3}}$, respectively. The best-fit ratio of CR to ram pressure $\xi_\mathrm{CR}$ for each model is listed in the last column of Table \ref{tab:fits}, when it is assumed that the gas density $\rho_2$ of shocked wind cavity is given by Eq. \eqref{eq:rho-cavity}. In contrast, when the gas density $\rho_2$ is a free parameter and its best-fit value for each model is listed in the second column of Table \ref{tab:fits-2}, we have assumed that $\xi_\mathrm{CR}=0.1$. The spatial diffusion coefficient $D_0=3\times 10^{25}~\mathrm{cm^2/s}$, while other parameters are specified in Section \ref{sec:sb}. The data points of HESS J1646-458 are taken from \citet{HC2022}.}
    \label{fig:gam-spec}
\end{figure}

The top panel of Figure \ref{fig:gam-spec} shows the $\gamma$-ray fluxes for three models with the external ISM mass density $\rho_0 =$ 30, 40, and 50 $m_p~\mathrm{cm^{-3}}$, respectively. The $\gamma$-ray flux is calculated according to Eq. \eqref{eq:gam-flux} once the distribution function of CR protons is got by solving Eq. \eqref{eq:parker}. Assuming that the mass density $\rho_2$ (the second column of Table \ref{tab:fits}) in the shocked wind cavity is given by Eq. \eqref{eq:rho-cavity}, the best-fit injection efficiency $\eta_\mathrm{inj}$ is listed in the second to last column of Table \ref{tab:fits}, while the third, fourth, and fifth columns give $R_\mathrm{TS}$, $R_S$, and the cutoff momentum $p_c$, respectively, for the three models. Furthermore, in the last column, the ratio of CR to ram pressure \footnote{\citet{Blasi2023} also call $\xi_\mathrm{CR}$ the efficiency of CR acceleration.}, $\xi_\mathrm{CR} = P_\mathrm{CR}/P_\mathrm{ram}$, where $P_\mathrm{ram} = \Lw/(2\pi R_\mathrm{TS}^2 \uw)$ and $P_\mathrm{CR} = \frac{4\pi}{3}\int_{p_\mathrm{inj}}^\infty E_\mathrm{k}(p) f_\mathrm{TS}(p) p^2 dp$, at the TS is listed. We find that the $\gamma$-ray emission from HESS J1646-458 can be well explained with hadronic models, however, the required $\xi_\mathrm{CR}$ is larger than 1 for all three models. Since the flux of CRs escaped from the SB \citep{Morlino2021} is

\begin{equation} \label{eq:fluxesc}
  \begin{aligned}
   \phi_\mathrm{esc}(p) &= \left(uf - D\frac{\partial f}{\partial r}\right)_{r=R_S} \\ &= \frac{\uw}{4} \frac{R_\mathrm{TS}^2}{R_S^2} \frac{f_\mathrm{TS}(p)}{1 - \exp\left[-\frac{\uw R_\mathrm{TS}}{4D(p)} (1 - R_\mathrm{TS}/R_S)\right]} \\ &\simeq f_\mathrm{TS}(p) \frac{\uw}{4} \frac{R_\mathrm{TS}^2}{R_S^2},
  \end{aligned}
\end{equation}

\noindent the escaped CR power $L_\mathrm{CR,esc} = 4\pi R_S^2 \int_{p_\mathrm{inj}}^\infty \phi_{\mathrm{esc}}(p) E_\mathrm{k}(p) 4\pi p^2 dp \simeq  \frac{3}{2} \xi_\mathrm{CR} \Lw$. Consequently, the escaped CR power is larger than the kinetic power of cluster wind for all above three models, and the hadronic scenario seems not to explain the $\gamma$-ray emission of HESS J1646-458.

However, it is possible that the average gas density of shocked wind cavity can be larger than that given by Eq. \eqref{eq:rho-cavity}, if there are some dense clumpy gases within the SB \citep{Blasi2023}. In fact, \citet{HC2022} find that the total gas hydrogen density traced by HI and CO in the region around the Wd 1 is about 13.7 $\mathrm{cm^{-3}}$, and the gas distribution traced by CO is not homogeneous, but rather clumpy. Therefore, we consider the alternative scenario in which the average gas density $\rho_2$ of shocked wind cavity is a free parameter. On the other hand, we assume that $\xi_\mathrm{CR} = 0.1$ as a quite reasonable value. The second column of Table \ref{tab:fits-2} shows the best-fit gas density $\rho_2$ for models with $\rho_0 =$ 30, 40, and 50 $m_p\ \mathrm{cm}^{-3}$, while the second to last column lists the corresponding injection efficiency $\eta_\mathrm{inj}$. We find that the hadronic scenario with an average gas density of about 1 $m_p\ \mathrm{cm}^{-3}$ within the shocked wind cavity can explain satisfactorily the $\gamma$-ray emission of HESS J1646-458. Since the $\gamma$-ray flux is proportional to the product $\xi_\mathrm{CR} \rho_2$, the best-fit $\gamma$-ray flux is the same as the corresponding one obtained assuming that the gas density $\rho_2$ is given by Eq. \eqref{eq:rho-cavity} for the three models with $\rho_0 =$ 30, 40, and 50 $m_p\ \mathrm{cm}^{-3}$, respectively, discussed in the previous paragraph. An average gas density of about 1 $m_p\ \mathrm{cm}^{-3}$ is quite likely, if there are dense clumpy gases within the shocked wind cavity. \citet{Blasi2023} assumed even a larger average gas density of about 10  $\mathrm{cm}^{-3}$ for explaining the $\gamma$-ray emission of Cygnus bubble, while \citet{Menchiari2024} analyzed recently the gas distribution within Cygnus bubble from observations and found the average gas hydrogen density is about 8.5 $\mathrm{cm}^{-3}$ (see their Fig. 3). If the Wd 1 is also the case similar to Cygnus bubble, then the required $\xi_\mathrm{CR}$ would be much smaller. Concludingly, under such conditions, the $\gamma$-ray emission of HESS J1646-458 is likely to be of hadronic origin.

The bottom panel of Figure \ref{fig:gam-spec} shows the radial profiles of $\gamma$-ray intensities above the threshold of 0.37 TeV for the three models with external ISM density $\rho_0$ = 30, 40, and 50 $m_p~\mathrm{cm}^{-3}$. The $\gamma$-ray intensity is calculated according to Eq. \eqref{eq:gam-intensity}. For comparing with the $\gamma$-ray radial profile of HESS J1646-458 \citep{HC2022}, we normalize our model $\gamma$-ray intensity to the peak data point of HESS J1646-458 which has the highest excess. It is evident that our models can explain reasonably the $\gamma$-ray radial profile of HESS J1646-458. Therefore, it is a strong evidence that the TS around the Wd 1 is the site where CRs are accelerated up to sub-PeV energies.

It is noted from the bottom panel of Figure \ref{fig:gam-spec} that the angular extensions (= $R_S/d$) for the models with $\rho_0 =$ 40 and 50 $m_p~\mathrm{cm^{-3}}$ are a bit smaller than that of HESS J1646-458, for a nominal distance $d$ = 3.9 kpc to the Wd 1. The angular extension of the SB is smaller for larger external ISM mass density $\rho_0$, thus $\rho_0$ larger than 50 $m_p~\mathrm{cm^{-3}}$ is not preferred by the present data. On the other hand, the angular extension for the model with $\rho_0 = 30~m_p~\mathrm{cm^{-3}}$ is almost same as that of HESS J1646-458. However, the model with $\rho_0$ smaller than 30 $m_p~\mathrm{cm^{-3}}$ will fit worse to the $\gamma$-ray radial profile. Given that we do know exactly the position of supershell and that the gas distribution within the shocked wind cavity could be nonuniform, we will not address this problem further, though the contribution from electrons can not be excluded presently.

\begin{figure}
    \centering
    \includegraphics[width=0.5\linewidth]{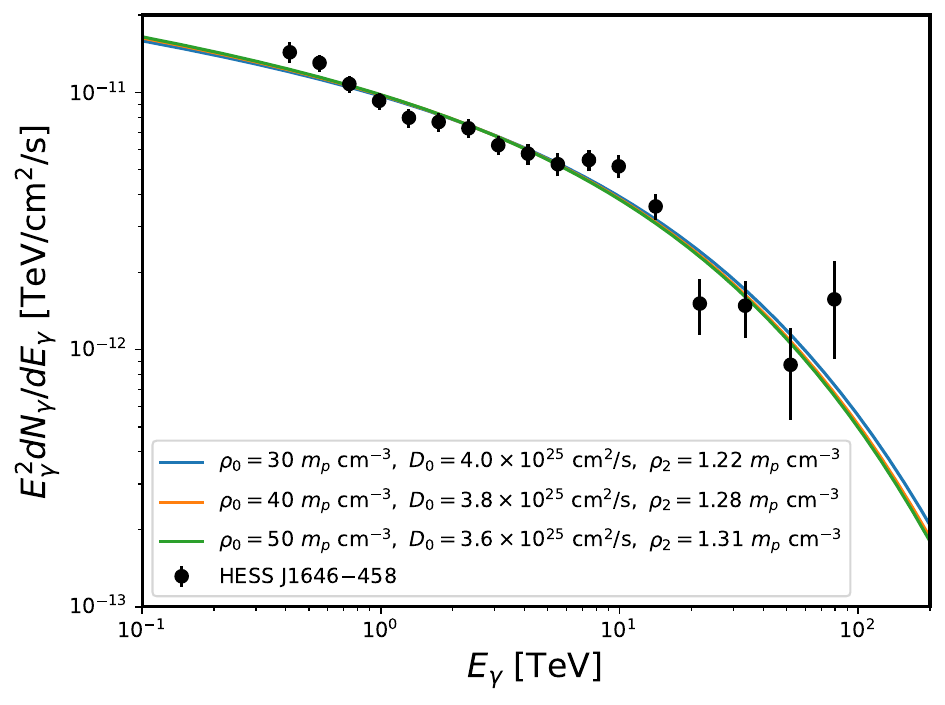}
    \caption{The $\gamma$-ray fluxes for three different models, which are got by fine-tuning the diffusion coefficient $D_0$, with the external ISM mass density $\rho_0$ = 30, 40, and 50 $m_p~\mathrm{cm^{-3}}$, respectively. For all the three models, the ratio of CR to ram pressure, $\xi_\mathrm{CR}=0.1$, is kept fixed. The best-fit diffusion coefficient $D_0$ = $4.0\times10^{25}$, $3.8\times10^{25}$, and $3.6\times10^{25}$ $\mathrm{cm^2/s}$, and the best-fit shocked wind cavity density $\rho_2$ = 1.22, 1.28, and 1.31 $m_p~\mathrm{cm^{-3}}$, respectively, for $\rho_0$ = 30, 40, and 50 $m_p~\mathrm{cm^{-3}}$, as shown explicitly in the legend. All other parameters are specified in Section \ref{sec:sb}. The data points of HESS J1646-458 are taken from \citet{HC2022}.}
    \label{fig:finetuning}
\end{figure}

Finally, we briefly mention the goodness of fits. Assuming that the ratio of CR to ram pressure $\xi_\mathrm{CR}$ = 0.1 is kept fixed, the reduced $\chi^2_\mathrm{red}$ is 3.6, 2.8, and 2.3 for models with external ISM mass density $\rho_0$ = 30, 40 and 50 $m_p~\mathrm{cm^{-3}}$, respectively, while the uncertainties of mass density $\rho_2$ are given within the parentheses in the second column of Table \ref{tab:fits-2}. By scanning the parameter space and especially fine-tuning the diffusion coefficient $D_0$, we find that the goodness of fit can be improved further. Figure \ref{fig:finetuning} shows the fits to the $\gamma$-ray emission of HESS J1646-458, which are obtained by fine-tuning the diffusion coefficient $D_0$, for models with $\rho_0$ = 30, 40, and 50 $m_p~\mathrm{cm^{-3}}$, respectively. The best-fit diffusion coefficient $D_0$ = $4.0\times10^{25}$, $3.8\times10^{25}$, and $3.6\times10^{25}$ $\mathrm{cm^2/s}$, and the best-fit shocked wind cavity density $\rho_2$ = 1.22, 1.28, and 1.31 $m_p~\mathrm{cm^{-3}}$, respectively, for $\rho_0$ = 30, 40, and 50 $m_p~\mathrm{cm^{-3}}$. Furthermore, the reduced $\chi_\mathrm{red}^2$ = 1.77, 1.74, and 1.74, respectively, for $\rho_0$ = 30, 40, and 50 $m_p~\mathrm{cm^{-3}}$. Therefore, while the $\gamma$-ray flux of HESS J1646-458 can be better modeled by fine-tuning transport parameters, our results are not influenced significantly.

\begin{table}[]
    \centering
    \begin{tabular}{ccccccc}
    \hline
    \hline
         $\rho_0$ & $\rho_2$ & $R_\mathrm{TS}$ & $R_S$ & $p_c$ & $\eta_\mathrm{inj}$ & $\xi_{\mathrm{CR}}$\\
         $[m_p~\mathrm{cm}^{-3}]$ & [$m_p~\mathrm{cm^{-3}}$] & [pc] & [pc] & [TeV/c] & [\%] & \\
    \hline
         30 & 0.058 & 25.5 & 96.0 & 281 & 9.61 & 1.62 \\
         40 & 0.067 & 23.4 & 90.6 & 217 & 9.21 & 1.52 \\
         50 & 0.076 & 21.8 & 86.7 & 177 & 8.93 & 1.45 \\
    \hline
    \end{tabular}
    \caption{Parameters for three best-fit models with the external ISM mass density $\rho_0 =$ 30, 40, and 50 $m_p~\mathrm{cm^{-3}}$, respectively, assuming that the mass density $\rho_2$ in the shocked wind cavity is given by Eq. \eqref{eq:rho-cavity} and the spatial diffusion coefficient $D_0=3\times 10^{25}~\mathrm{cm^2/s}$.}
    \label{tab:fits}
\end{table}

\begin{table}[]
    \centering
    \begin{tabular}{ccccccc}
    \hline
    \hline
         $\rho_0$ & $\rho_2$ & $R_\mathrm{TS}$ & $R_S$ & $p_c$ & $\eta_\mathrm{inj}$ & $\xi_{\mathrm{CR}}$\\
         $[m_p~\mathrm{cm}^{-3}]$ & [$m_p~\mathrm{cm^{-3}}$] & [pc] & [pc] & [TeV/c] & [\%] & \\
    \hline
         30 & 0.934 (0.023) & 25.5 & 96.0 & 281 & 0.59 & 0.1 \\
         40 & 1.023 (0.025) & 23.4 & 90.6 & 217 & 0.61 & 0.1 \\
         50 & 1.101 (0.027) & 21.8 & 86.7 & 177 & 0.62 & 0.1 \\
    \hline
    \end{tabular}
    \caption{Parameters for three best-fit models with the external ISM mass density $\rho_0 =$ 30, 40, and 50 $m_p~\mathrm{cm^{-3}}$, respectively, assuming that the ratio of CR to ram pressure $\xi_\mathrm{CR} = 0.1$ and the spatial diffusion coefficient $D_0 = 3\times 10^{25}~\mathrm{cm^2/s}$. The values within the parentheses in the second column are the uncertainties of best-fit mass density of shocked wind cavity. The values of $R_\mathrm{TS}$, $R_S$, and $p_c$ are the same as those in Table \ref{tab:fits}.}
    \label{tab:fits-2}
\end{table}

\section{Conclusion and discussion} \label{sec:conclusion}

In the present work, we developed the model of CR acceleration at the TS forming around the Wd 1, and calculated the spectrum and radial profile of $\gamma$ rays produced from the inelastic collisions of hadronic CRs with the ambient gas. With the reasonable selection of parameters, our results can fit well to the spectrum and radial profile of the $\gamma$-ray emission of HESS J1646-458. When the gas density within the shocked wind cavity is given by Eq. \eqref{eq:rho-cavity} as predicted by the classical wind-blown bubble model \citep{MacLow1988}, the ratio of CR to ram pressure $\xi_\mathrm{CR}$ is required to be larger than 1, thus the hadronic scenario is not preferred by the present observational data. However, if there are dense clumpy gases in the shocked wind cavity \citep{Blasi2023} such that the average gas density is much larger than that given by Eq. \eqref{eq:rho-cavity}, then the $\gamma$-ray emission of HESS J1646-458 can be explained satisfactorily. In particular, if the average gas density in the shocked wind cavity is about 1 $m_p~\mathrm{cm}^{-3}$, the required ratio of CR to ram pressure $\xi_\mathrm{CR}$ is about 0.1, which is a reasonable value. As a consequence, the $\gamma$-ray emission of HESS J1646-458 is likely to be of hadronic origin, demonstrating that the TS is a capable accelerator of CRs. Therefore, the Galactic YMCs can have an unignorable contribution to the Galactic CRs up to sub-PeV energies.

We have investigated the acceleration of CRs at the TS in the test particle approximation. Under such an approximation, the back-action of CR pressure on the TS is not taken into account. When the CR pressure is not negligible compared to the ram pressure at the TS, CRs can modify the background fluid flow \citep{Malkov2001}. Moreover, the self-generated turbulence driven by the streaming instability due to the CR gradient can have significant effects on the acceleration and confinement of CRs near the TS \citep{Marcowith2021}. However, we will not further discuss nonlinear effects on the acceleration of CRs, whereas it is beyond the scope of the present article. These nonlinear effects should be incorporated in a more complete model for addressing the acceleration of CRs by YMCs in more detail in future.

The supershell has a very high gas density as implied by Eq. \eqref{eq:mass}, thus it should be an ideal target for CRs. However, no any structure related to the supershell in the $\gamma$-ray emission of HESS J1646-458 is detected by H.E.S.S. \citep{HC2022}, while the shell-like structure in the $\gamma$-ray emission can not be attributed to the supershell, since the required external ISM density will be significantly larger than the typical values otherwise. We have assumed that the supershell is infinitely thin, and the absorbing boundary condition is adopted, consequently the transport of CRs in the supershell can not be addressed in our model. For addressing the transport of CRs in the supershell, a more realistic model of SBs should be adopted. 

For accelerating CRs effectively, we got a quite small diffusion coefficient $D_0 = 3\times 10^{25}~\mathrm{cm^2/s}$ inside the SB, which is 3 orders of magnitude lower than that in the diffuse ISM \citep{Strong2007}. It is generally believed that the Bohm diffusion coefficient $D_\mathrm{Bohm} = \beta pc^2/3ZeB = 10^{23} \beta Z^{-1} (B / 1~\mu\mathrm{G})^{-1} (p / 3~\mathrm{GeV~c^{-1}})~\mathrm{cm^2/s}$, where $Z$ is the charge number, $e$ is the elementary charge, and $B$ is the magnetic field, is the smallest accessible diffusion coefficient for CRs, then a magnetic field of at least 11 $\mathrm{\mu G}$ is needed if Eq. \eqref{eq:diffcoeff} is applicable up to 1 PeV for the Kolmogorov turbulence spectrum. Such a magnetic field is larger than the upper bound of 7 $\mathrm{\mu G}$ given recently by the eROSITA X-ray observation to the Wd 1 \citep{Haubner2025}. Consequently, the observation of eROSITA appears to be unfavorable to the hadronic scenario. Such a discrepancy, however, is not so serious, considering the uncertainty in inferring the magnetic field via the eROSITA. It should be noted that the eROSITA upper bound of magnetic field is obtained by assuming the $\gamma$-ray emission of HESS J1646-458 is purely leptonic. For leptonic scenario, the electron energy should be up to at least 100 TeV for producing several tens of TeV $\gamma$-ray emission. Electrons with energies of about 100 TeV should also produce synchrotron X-ray emission, which is constrained by the upper bound of synchrotron flux estimated by the eROSITA \citep{Haubner2025}. However, according to Figure \ref{fig:espec}, we have seen that it is hard for electrons to be accelerated up to 100 TeV due to severe energy losses resulting from synchrotron radiation and IC scattering off background soft photons. On the other hand, if the $\gamma$-ray emission of HESS J1646-458 is predominantly of hadronic origin, then a magnetic field larger than 7 $\mathrm{\mu G}$ should be possible. Therefore, it is hard with the present observational data to pin down the dominant radiation process contributing to the $\gamma$-ray emission of HESS J1646-458, and the hadronic scenario can not be excluded as claimed by \citet{Haerer2023}. Future multiwavelength and multimessenger observations to the Wd 1 should give more clues and place more stringent constrains on the nature of the $\gamma$-ray emission of HESS J1646-458.

\section*{Acknowledgments}
R.-Z. Yang is supported by the National Natural Science Foundation of China under grants 12588101, 12393854, and by the natural science funding of Sichuan Province under grant 2025ZNSFSC0065. R.-Z. Yang gratefully acknowledges the support of Cyrus Chun Ying Tang Foundations and of the studio of Academician Zhao Zhengguo, Deep Space Exploration Laboratory.

  \bibliographystyle{elsarticle-harv} 
  \bibliography{main.bib}





\end{document}